\documentclass[twocolumn]{revtex4-1}
\usepackage{graphicx} 

\begin{document}
\title{The Gravity Tunnel Superhighway}
\author{Alexander R. Klotz}
\affiliation{Department of Physics and Astronomy, California State University, Long Beach}

\begin{abstract}
This manuscript discusses gravity tunnels formed by connecting two vertical shafts by a constant-radius tunnel within the Earth, which featured in a dream I had in September 2024. The total travel time through such a tunnel can be minimized with respect to the radius at which the shafts are connected. I derive this minimal radius and minimum time given two assumptions for Earth's interior, that of constant gravitational acceleration and that of uniform density. Both models have solutions in terms of basic functions, and are typically 10\% slower than the brachistochrone curve between the same points. I also find the optimal depth of a ``superhighway,'' which minimizes the average time to fall between any two points on a great circle. Finally, I discuss the role of problems like these in physics education.

\end{abstract}

\maketitle

\section{Introduction}

This idea for this manuscript came to me in a dream and concerns simplified gravity tunnels through the Earth that consist of two straight radial shafts connected by an angular segment at constant radius (Figure 1). A traveler would fall straight down building up speed, slide along the angular portion at constant speed after a smoothed transition, and then decelerate up towards the destination, reaching it at zero velocity. Finding the optimal way to connect two points on the Earth by such a tunnel may serve as a useful way to introduce the idea of minimization to undergraduate physics students, without the more complex ideas that typically come along with it, such as functionals, differential geometry, the Euler-Lagrange equation, and tricks to solve ordinary differential equations. Gravity tunnel problems are typically solved assuming a uniform density Earth, which leads to simple harmonic motion. In my 2015 paper ``The Gravity Tunnel in a non-Uniform Earth,'' \cite{klotz} I showed that a simpler assumption that the gravitational field strength is constant yields results much closer to those found from Earth's actual density profile. In this work, I will discuss the derivation of minimal-time solutions to this particular gravity tunnel problem according to both assumptions, as well as how they may be incorporated into an undergraduate physics education.

\begin{figure}[h]
    \centering
    \includegraphics[width=0.5\linewidth]{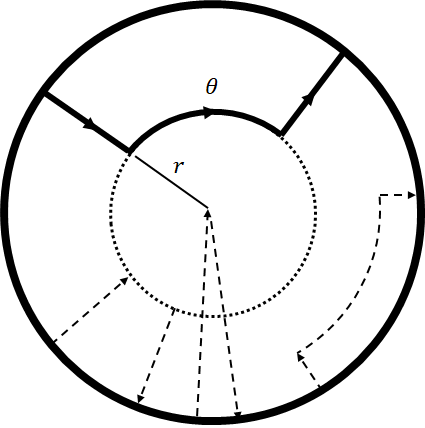}
    \caption{Schematic of tunnels through a planet that connect two vertical shafts separated by angle $\theta$ with an angular component at constant radius $r$. The thickest path shows a potentially optimized route, while others are likely not optimized. The dashed inner circle represents a ``superhighway'' that can be reached by multiple shafts from the surface.}
    \label{fig:enter-label}
\end{figure}
The typical gravity tunnel problem begins by asking how long it would take to fall through a tunnel from one point on the Earth through its center to an antipodal point. Ignoring factors such as rotation, friction, and engineering, the assumption that the Earth is uniformly dense leads to a gravitational field that is linear with radial position. The linear force implies simple harmonic motion, and the time may be determined by the half-period of oscillation, 42 minutes, a solution known for hundreds of years \cite{redier}. The next level of the problem involves connecting any two non-antipodal points with a straight tunnel and showing that the same simple harmonic motion arises with the same period, and that any  two points may be traversed in the same 42 minutes \cite{cooper}. Finally, one may ask to find the brachistochrone path which connects any two points in the shortest time. The solution is that of a hypocycloid curve, produced by a circle rolling inside another \cite{venezian}. In the constant gravity assumption, it is easier to derive the antipodal fall time of 38 minutes, but the line time is no longer independent of surface distance, and both it and the brachistochrone path have much more complex solutions in terms of elliptic integrals \cite{klotz}.

If, instead of finding the fastest path or the shortest path, we wish to find a fast but simple path, we can dig straight down from the two termini, and connect the two shafts with another tunnel segment at constant radius. The dynamics in the vertical section are that of freefall, the angular section is that of constant-velocity angular motion, so the travel time can be calculated simply. The challenge is determining at which radius the connecting segment should be built do minimize the total time. One could then imagine, in a distant future, a single circular tunnel dug at some radius within the Earth, that various cities sink shafts towards to connect themselves to this gravitational superhighway. At what radius should this common tunnel be dug?

What considerations should be taken into account to minimize the time? The travel time is the ratio of distance to average speed, so distance should be small and speed large. A deep tunnel will maximize speed over a longer distance, while a short tunnel will minimize distance over a slower trip. Before doing any calculations, one can consider that radius of the sliding tunnel cannot be less than zero. Long trips with a very small sliding radius should be compared to the case of simply digging two shafts that meet at Earth's center. For any two points separated by more than 2 radians (12,760 km, approximately the distance between Los Angeles and Melbourne), falling down to the center and then up to the destination will be faster than two shafts connected by a sliding component. We also know that the fastest curve connecting any two points is the brachistochrone, and the shaft-and-slide tunnel will always be slower.


\section{Constant Gravity}
Earth's gravitational acceleration can be treated as constant for gravity tunnel problems because the increasing density towards the core is such that the gravitational field strength weakly increases through the mantle, and comparatively little time is spent falling through the core where the field varies more. The time $T$ it takes to traverse such a tunnel is the time it takes to fall down the shaft, $T_f$, the time it takes to slide along the constant radius segment, $T_s$, and the time it takes to rise up the final shaft, which by symmetry is also $T_f$:
\begin{equation}
    T=2T_{f}+T_{s}.
\end{equation}

If the shaft goes from the surface at radius $R$ to some smaller radius $r$, covering distance $R-r$, the time it takes to fall from rest under constant acceleration $g$ can be found from simple kinematics:
\begin{equation}
    R-r=\frac{1}{2}gT_{f}^2\rightarrow T_{f}=\sqrt{2\frac{(R-r)}{g}}.
\end{equation}
The velocity reached at the bottom can be found from energy conservation, assuming rest at the surface:
\begin{equation}
    \frac{1}{2}mv^{2}+mgr=mgR\rightarrow v=\sqrt{2g(R-r)}.
\end{equation}
If the sliding phase covers the angle $\theta$ that separates the two surface points, this yields a constant gravity fall time $T_g$ of:
\begin{equation}
    T=2\sqrt{2\frac{(R-r)}{g}}+\frac{r\theta}{\sqrt{2g(R-r)}}.
\end{equation}
To find the minimal time we differentiate with respect to $r$ and find its zero:
\begin{equation}
    \frac{dT}{dr}=\frac{1}{\sqrt{8}}\left[\frac{-4}{\sqrt{g(r-R)}}+\frac{gr\theta}{(g(R-r)^{\frac{3}{2}}}+\frac{2\theta}{\sqrt{g(R-r)}}\right]=0
\end{equation}
The radius for minimal time, $r_g$ is:
\begin{equation}
    r_{g}=2R\frac{\theta-2}{\theta-4}.
\end{equation}
This is substituted back into the expression for $T$ to find $T_{g}$:
\begin{equation}
    T_{g}=\sqrt{\frac{R}{g}(8\theta-2\theta^2)}
\end{equation}
As expected, $r_g$ approaches zero as the angle traversed approaches 2 radians, and $T_g$ approaches the diameter fall time $\sqrt{8\frac{R}{g}}$ in the same limit, 38 minutes for the Earth. This can be compared to the brachistochrone curve within the Earth, which for the constant gravity assumption is extremely complicated. We can approximate the time in minutes with a fourth-order polynomial for $\theta$ between 0 and 2:
\begin{equation}
    T_{gb}\approx-4.4\theta^{4}+22.7\theta^{3}-45.9\theta^{2}+51.4\theta+5.4.
\end{equation}
This simply allows us to plot it more easily against $T_g$ but does not convey additional insight about the brachistochrone curve in constant radial gravity, and clearly fails for small angles. By analogy, we can also compare to the problem of falling down, over, and up in a vertical field in Cartesian geometry and compare that to Bernoulli's cycloid brachistochrone. This is also a suitable question for an undergraduate, the optimal height to fall in order to reach distance D in the least time is D/4, and it takes $T=\sqrt{8D/g}$. This is a factor of $4/\pi\approx 1.27$ deeper than the brachistochrone, and a factor of $\sqrt{4/\pi}\approx 1.12$ slower. 
\section{Uniform Density}
The typical gravity tunnel problem assumes a uniformly dense Earth leading to simple harmonic motion through the tunnel. Although this uses more complex physics to achieve a less accurate answer compared to the constant gravity assumption, the straight tunnel and brachistochrone solutions are much simpler. It is also useful in conveying to introductory physics students that simple harmonic motion can apply to things besides masses on springs, an important lesson for later in their journeys.

For a vertical shaft, we can set up an equation of motion based on Newton's law of gravity, with a gravitating mass increasing with the cube of radial position due to the shell theorem:
\begin{equation}
    F=m\ddot{r}=-\frac{G\rho (\frac{4}{3}\pi r^{3})m}{r^2}=-mg\frac{r}{R}\rightarrow \ddot{r}+\frac{g}{R}r=0.
\end{equation}
The simplification above is based on replacing the Newtonian gravitational terms with the definition of $g$. This reduces to the familiar form of the simple harmonic equation of motion, which has a solution in terms of sines and cosines with an angular frequency given by the square root of the linear coefficient. Starting at rest at the surface, we can write the solution as:
\begin{equation}
    r(t)=R\cos\left(\sqrt{\frac{g}{R}}t\right) \ \ \ \& \ \ \ v(t)=-\sqrt{gR}\sin\left(\sqrt{\frac{g}{R}}t\right). 
\end{equation}
To find the time taken for a falling object to reach a specific R, we need to invert the equation of motion to find $T_{f}(r)$:
\begin{equation}
    T_{f}(r)=\sqrt{\frac{R}{g}}\arccos\left(\frac{r}{R}\right).
\end{equation}
Since the shaft in question does not reach the center, the phase is between 0 and $\pi/2$ and there are no issues with ambiguities of inverse trigonometric functions. The speed that is reached for the sliding portion can be found by substituting $T_f$ into the velocity equation:
\begin{equation}
    v=\sqrt{gR}\sin\left(\sqrt{\frac{g}{R}}\sqrt{\frac{R}{g}}\arccos\left(\frac{r}{R}\right)\right)=\sqrt{gR\left(1-\left(\frac{r}{R}\right)^{2}\right)}
\end{equation}
The sine of an inverse cosine is simply a switch from the horizontal to the vertical component of the unit circle (and vice versa), and can be substituted with the Pythagorean theorem. The total travel time is then:
\begin{equation}
    T=2\sqrt{\frac{R}{g}}\arccos\left(\frac{r}{R}\right)+\frac{r\theta}{\sqrt{gR\left(1-\left(\frac{r}{R}\right)^{2}\right)}}
\end{equation}
We can follow the same minimization procedures:
\begin{equation}
    \frac{dT}{dr}=\frac{1}{\sqrt{gR}}\left[\frac{2+\theta}{\sqrt{1-\frac{r^2}{R^2}}}+\frac{r^2}{R^2}\frac{\theta}{\left(1-\frac{r^2}{R^2}\right)^{3/2}}\right]=0
\end{equation}
Equations such as (5) and (14) typically have two solutions, one of which is unphysical and has $r>R$. Taking the physical solution, the radius for minimal time is:
\begin{equation}
    r_{\rho}=R\sqrt{1-\frac{\theta}{2}}
\end{equation}
The minimal time under the uniform density assumption is then:
\begin{equation}
    T_{\rho}=\sqrt{\frac{R}{g}}\left[\sqrt{2\theta-\theta^{2}}+2\arcsin{\sqrt{\frac{\theta}{2}}}\right]
\end{equation}
For comparison, the hypocycloid brachistochrone is traversed in time:
\begin{equation}
    T_{g,\rho}=\sqrt{\frac{R}{g}}\sqrt{\theta(2\pi-\theta)}
\end{equation}
Figure 2 shows, as a function of surface angle $\theta$, the minimum depths according to the two models, and the minimum times in comparison to the brachistochrones. Both solutions are about 10\% slower than the brachistochrone, the ratio of times decreasing with angular separation, consistent with the 12\% slowdown for the vertical analogue of this problem.

\begin{figure*}
    \centering
    \includegraphics[width=1\linewidth]{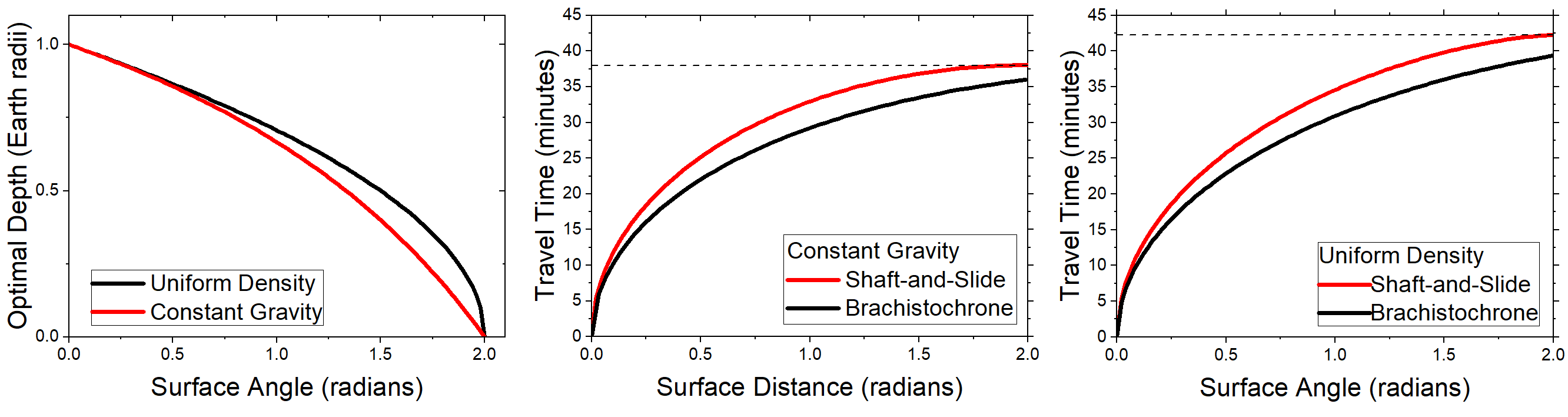}
    \caption{Left. Optimal radius of the angular portion of the tunnel to minimize travel time, as a function of the angle traversed. Middle. Minimum travel time along shaft-and-slide paths as a function of the angle traversed for the constant gravity model, with the brachistochrone time shown for comparison. Right. Minimum travel time along shaft-and-slide paths as a function of the angle traversed for the uniform model, with the brachistochrone time shown for comparison. Dashed lines represent the diameter time.}
    \label{fig:times}
\end{figure*}

\section{The Gravity Tunnel Superhighway}
The year is 3024 and the governments of the world representing the united peoples of Earth decide to build a common transportation network based on a single circular tunnel running below a great circle on the surface. Each major city may sink a shaft down and connect to this tunnel, allowing rapid fuel-free transportation between the world's population centers. Where should this circular tunnel be placed? Directly under the equator it would most readily service cities such as Singapore, Kuala Lumpur, Quito, Belem, Kinshasa, and Nairobi, with cities farther from the equator digging longer shafts to reach it. A tunnel in a perpendicular plane running beneath the 90th meridians would service much of the Eastern United States, Latin America, India, Southeast Asia, and Western China. I am not aware of a great circle that maximizes the population living near it, but finding an ideal one is a challenge as most of the planet's surface is ocean. Finding the depth of such a tunnel is more answerable.
For a given radius $r$, the travel time averaged over every possible angular separation can be written as a weighted average of $T(r,\theta)$ with $\theta$ as the weighting:
\begin{equation}
\langle T(r)\rangle=\frac{\int_0^\pi T(r,\theta)d\theta}{\int_0^\pi d\theta}=\frac{1}{\pi}\int_0^\pi T(r,\theta)d\theta     
\end{equation}
Here, $T$ may be substituted with either the constant gravity or uniform density equations above. Since both are only linear functions of $\theta$ the integration is not complicated, the constant term picks up a factor of $\pi$ and the linear term a factor of $\pi^2/2$. To find the radius that minimizes the average travel time, we differentiate $\langle T\rangle$ and find $r_{av}$ such that $d\langle T\rangle/dr=0$. Doing so we find the following values:
\begin{equation}
    r_{av,g}=2R\frac{4-\pi}{8-\pi}\approx0.35R
\end{equation}
and
\begin{equation}
    r_{av,\rho}=\frac{R}{2}\sqrt{4-\pi}\approx0.46R.
\end{equation}
Both of these radii are within the outer core.
\section{Role in Physics Education}
One of the central principles of advanced physics is the principle of least action, although it is typically not introduced until the third or fourth year of an undergraduate physics degree. There have been initiatives to introduce it earlier in the curriculum, as an extension of Newton's laws or energy conservation \cite{taylor2}. Students often learn about tools to minimize the action, such as the Euler-Lagrange equation, at the same time as the action or Lagrangian is introduced. There is a risk of students internalizing Lagrangian or Hamiltonian mechanics as mere tools that turn potentials into equations of motion, without appreciating the importance of the underlying principle. Earlier on in a students education, there are action-related concepts that are introduced that can seem magical or ascribe agency to fundamental particles, such as Fermat's principle that light always follows the path of shortest time, or that electricity always follows the path of least resistance, and these can lead to lingering misconceptions.
The brachistochrone problem is introduced both for historical reasons, as its solution lead to the development of variational calculus, and for conceptual reasons, as it is easier to understand minimizing time than minimizing the time integral of the difference of kinetic and potential energy. Taylor's Classical Mechanics \cite{taylor}, for example, has the brachistochrone as the second example after the Euler-Lagrange equation is derived and before the Lagrangian is introduced. However, as stated in the introduction, the significance can be lost if too many concepts are introduced at once. Depending on their background, the discussion of the brachistochrone problem may be occurring around the first time they are seeing a functional, an arc length differential, the Euler-Lagrange equation, and the differential equation techniques used to solve it. It is likely something the instructor does on the board, rather than something they do themselves.
There is a benefit to introducing minimization and optimization concepts that can be solved by students using tools that they already know. These can be time-minimization problems with arbitrary paths of known form, as in this manuscript. One example that appears in undergraduate education is the ``lifeguard'' derivation of Snell's law from Fermat's principle. It may be worthwhile assigning problems such as these in introductory mechanics or early in an advanced mechanics class.  Students can become familiar with the idea of finding a minimal trajectory for a problem, without requiring new mathematical techniques at the same time that can obfuscate the physics. Some examples include the Snell's lifeguard problem, minimizing the down-over-up sliding time, minimizing the time to fall along two connected line segments, and the gravity tunnel problem in this manuscript. Beyond minimizing time, one may devise problems in which the action is computed for set trajectories, showing that it is only minimized for trajectories which solve an equation of motion.

\section{Concluding Remarks}
After my paper on gravity tunnels was published, a flurry of papers were published examining the effects of friction \cite{concannon2016gravity}, rotation \cite{isermann2019free}, the shape of the Earth \cite{taillet2018free}, Keplerian orbits \cite{gjerlov2019orbits}, and even relativity \cite{seel2018relativistic} (although many of these concepts were discussed by Andrew Simoson before me \cite{simoson2007hesiod}). By contrast, the brachistochrone inside a rotating gravitating body has not been explored in as much depth, with the exception of Crosswhite and Antman who showed that brachistochrone paths in the sagittal plane of a uniformly dense body may not have exits \cite{crosswhite2006new}. As the vertical brachistochrone is a cycloid and the uniform planetary brachistochrone is a hypocycloid, the brachistochrone between two points in a rotating plane is an epicycloid. Combining these various trochoids in the equatorial plane of a spinning planet may lead to some interesting results, but would likely be very hard. An easier problem, ripe for a keen solver, is the extension of the results of this manuscript to a rotating planet.

\section{Acknowledgements}
The author is grateful for his institution not holding classes the week of Thanksgiving, giving him time to write this document. An early version of the constant gravity minimization was worked out by Holland Karaghiaulleian.

\bibliographystyle{unsrt}
\bibliography{earthrefs}

\end{document}